\documentclass[twocolumn,aps,prl,superscriptaddress,showpacs,a4paper]{revtex4}
\usepackage{graphicx}
\usepackage{amsfonts,amssymb,amsmath,bbm}
\usepackage[utf8]{inputenc}

\begin{document}
\title{Speckle instability: coherent effects in nonlinear disordered media} 

\author{Beno\^{\i}t Gr\'emaud}
\affiliation{Laboratoire Kastler Brossel, Ecole Normale Sup\'erieure,
 CNRS, UPMC, 4 place Jussieu, 75252 Paris Cedex 05, France}
\affiliation{IPAL, CNRS, I2R,  1 Fusionopolis Way, South Tower, Singapore 138632}
\author{Thomas Wellens}
\affiliation{Physikalisches Institut, Albert-Ludwigs Universit\"at Freiburg, Herman-Herder-Str. 3a, 79104 Freiburg, Germany}
%\email{benoit.gremaud@spectro.jussieu.fr}
\date{\today}

\begin{abstract}
 We numerically investigate the properties of speckle patterns formed by nonlinear
 point scatterers. We show that, in
 the weak localization regime, dynamical
 instability appears, eventually leading to chaotic  behavior of
 the system. Analysing the statistical properties of the instability
 thresholds for different values of the system size and disorder
 strength, a scaling law is emphasized. The later is also found to
 govern the smallest decay rate of the linear system,
 putting thus forward the crucial importance of interference
 effects. This is also underlined by the fact that
 coherent backscattering is still
 observed even in the chaotic regime.
\end{abstract}


\maketitle

As first described by Anderson in 1958, the impact of disorder on the transport properties of waves, depending on the dimensionality and the disorder strength, ranges from weak to strong localization. In the case of matter waves, the localization of Bose-Einstein condensates (BEC)
is, at present, a very active research topic investigated by several experimental and theoretical groups~\cite{clement05,fort05,schulte05,paul07,robert07,lsp07,skipetrov08,hartung08}, in particular with the recent observation of the (1D) localization of matter waves in a disordered optical  potentials~\cite{lsp08,roati08}. Although, in these experiments, the atom-atom interactions were negligible, the basic question remains how 
effects of interference between multiply scattered waves, such as weak or strong localization,  are affected by interactions. 
 
BEC's in disordered potentials appear to be good candidates to study these questions, since, in the mean field regime, the condensate is still described by a single coherent wave function governed by a nonlinear wave equation (Gross-Pitaevski equation). Thus,
the condensate in principle retains its ability to display interference effects also in presence of (not too strong) interactions. Similar nonlinear equations describe propagation of light in disordered nonlinear media \cite{skipetrov}. In contrast, the situation is quite different in the
case of electronic transport \cite{altshuler99}, where the interactions combined with finite temperature effects give rise to dephasing, which in general destroy the disorder-induced coherent effects.

Even if the theoretical description of coherent effects in nonlinear disordered systems is far from complete,
an important step was done recently by the development of a diagrammatic theory for coherent backscattering in presence of nonlinearity \cite{wellens08}. This approach relies on the assumption of a unique stationary solution of the nonlinear wave equation under consideration. This assumption is expected to be valid for very weak nonlinearities
(which, however, may still considerably affect the height of the coherent backscattering cone \cite{wellens08,hartung08}).
 On the other hand, it is known that larger  nonlinearities 
 %the wave equation
%may admit a multiplicity of stationary solutions~\cite{spivak}. Furthermore, the nonlinearity 
can induce
speckle instabilities, such that no stationary state is reached at long times \cite{hartung08,paul05}. 
A clear explanation for the physical origin of this effect, however, is still missing.
Theoretical predictions of the nonlinearity
threshold above which instabilities develop, were attempted in
\cite{spivak,skipetrov}, but, to our knowledge, these results  have not yet been confirmed by experiments
or numerical simulations.
For this reason, we investigate, in this letter, the \textit{dynamical} properties of the speckle patterns generated by nonlinear point scatterers.

More precisely, we consider an assembly of $N$ point-like
scatterers located at randomly chosen 
positions $\mathbf{r}_i$, $i=1,\dots,N$, inside a sphere of volume $V$
illuminated by a plane wave $\mathbf{k}_L$. The field amplitude $E_i$ at scatterer $i$
results as the sum of the incident wave and the field radiated by all other scatterers:
\begin{equation}
\label{model}
E_i=e^{i\mathbf{k}_L\cdot \mathbf{r}_i}+\sum_{j\neq i}
\frac{e^{ik|{\bf r}_i-{\bf r}_j|}}{k|{\bf r}_i-{\bf r}_j|} d_j,
\end{equation}
where $k=|{\bf  k}_L|=2\pi/\lambda$, and the field amplitudes are measured in units of the incident plane wave
amplitude. $d_j$ describes
the dipole induced inside scatterer $j$. For simplicity, we will
consider only scalar fields in this letter. 

As for the induced dipole, we consider a nonlinear response $d=g\left(|E|^2\right)E$, with
\begin{equation}
\label{nlps}
g(I)=i \frac{e^{i\alpha I}+1}{2}.
\end{equation}
For $\alpha=0$, this model corresponds to a resonant point scatterer. The nonlinearity $\alpha$
induces a phase shift proportional to the local intensity $I$, which amounts to a shift of
the scatterer's resonance frequency.
For small $\alpha$, Eq.~(\ref{nlps}) reduces to a
$\chi^{(3)}$-nonlinearity; furthermore, for real values of $\alpha$,
the optical theorem is fulfilled, ensuring energy conservation.
The above set of $N$ complex equations \eqref{model}, with the replacement \eqref{nlps}, is formally written as $\mathbf{F}(\mathbf{E},\alpha)=\mathbf{0}$, where $\mathbf{F}$ and $\mathbf{E}$ are real vectors of dimension $2N$.

Starting from the linear solution ($\alpha=0$) for a given configuration of the scatterers, a numerical scheme, based on a Newton-Krylov method, allows to continuously
follow the solution $\mathbf{E}(\alpha)$, which fulfills
\begin{equation}
\frac{d\mathbf{E}}{d\alpha}=
-{\mathbf M}^{-1}\frac{\partial\mathbf{F}}{\partial\alpha}
\end{equation}
with  the Jacobian matrix $\mathbf{M}=\frac{\partial\mathbf{F}}{\partial\mathbf{E}}$.
If $\mathbf{M}$ has a vanishing eigenvalue, 
this gives rise to a turning point in the curve $\mathbf{E}(\alpha)$, such that several solutions
of Eq.~(\ref{model}) coexist for the same value of $\alpha$. 

\begin{figure}
\centerline{\includegraphics[width=6cm]{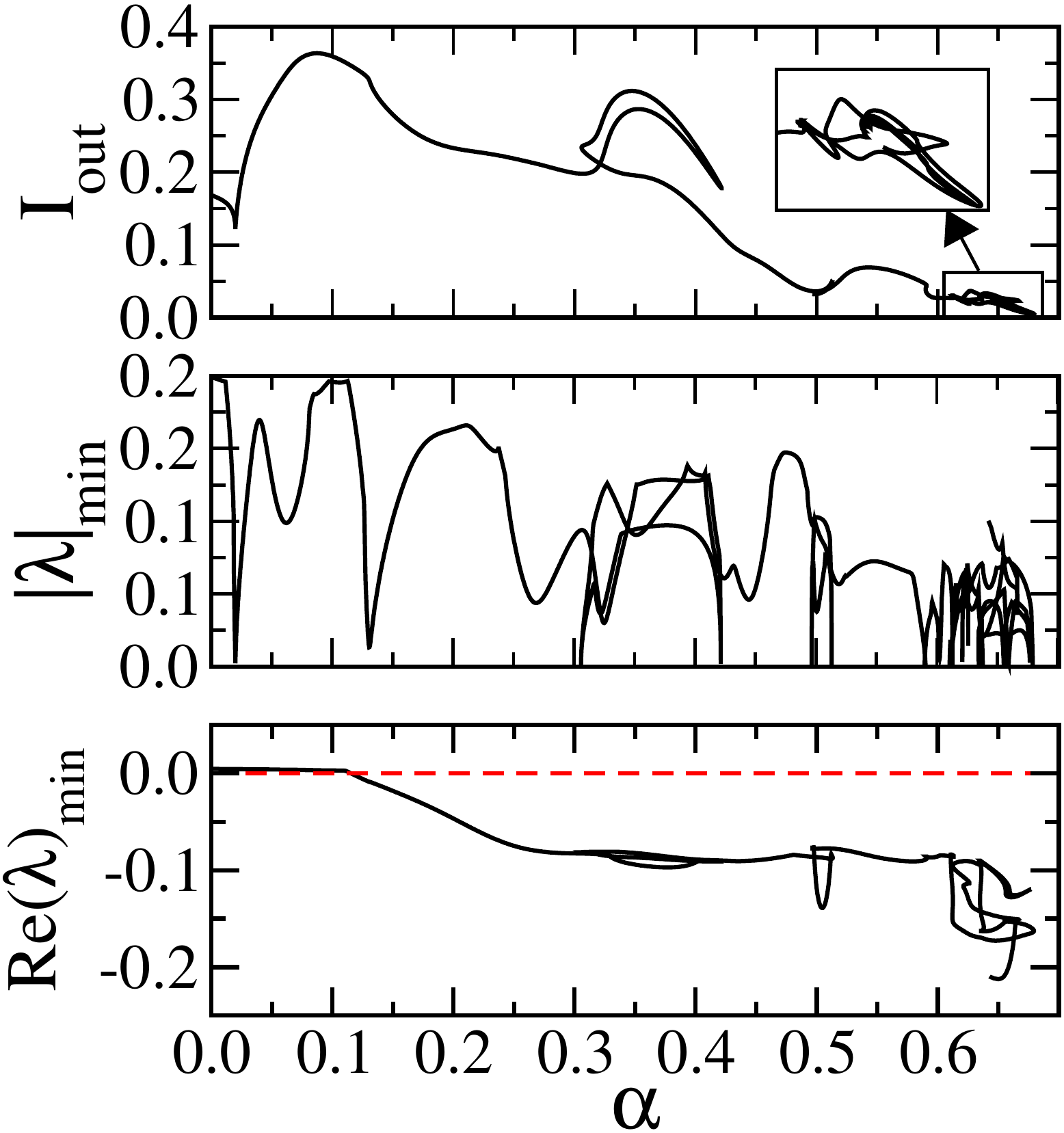}}
\caption{\label{multisol} (Color online) For a given random configuration of $N=1000$
scatterers with density $n\lambda^3=1$, the top plot depicts the backward
intensity $I_\mathrm{out}$ as a function of the nonlinearity strength
$\alpha$. For $\alpha>0.3$, several stationary solutions coexist.
The middle plot shows the smallest module of all
eigenvalues of the stability matrix $\mathbf{M}$. At each turning point, $\mathbf M$ exhibits a vanishing eigenvalue. The bottom plot
depicts the smallest real part of all
eigenvalues of $\mathbf M$. The dashed
line indicates the stability threshold defined by $\mathrm{Re}(\lambda)=0$. At $\alpha\simeq 0.1$, 
the solution becomes unstable, and
the instability then increases with $\alpha$.}
\end{figure}

As an example, we plot in Fig.~\ref{multisol} (top) the
intensity $I_\mathrm{out}$, scattered into the direction opposite to the incident wave 
as a function of $\alpha$, for a specific configuration
of $N=1000$ scatterers randomly distributed in a sphere with density $n=N/V$ such that $n\lambda^3=1$. 
%In the linear regime (
For $\alpha=0$, these parameters 
correspond to mean free path $\ell=20/k$ and 
optical thickness (measured along the diameter $2R$ of the sphere)
$2R/\ell_0=4$. In Fig.~\ref{multisol} (top), we clearly
notice the existence of several stationary solutions
for $\alpha>0.3$, and in particular for $\alpha>0.6$ (see inset).
%Furthermore, we observe that the backscattered intensity tends to decrease for
%increasing nonlinearity. This is explained by the decreasing cross section of each single scatterer 
%due to the shift of the resonance frequency induced by the nonlinearity.
Fig.~\ref{multisol}(middle) shows the smallest module $|\lambda|_{\mathrm{min}}$
of all eigenvalues of the matrix $\mathbf{M}$. At every turning point, the curve $|\lambda|_{\mathrm{min}}(\alpha)$ touches the horizontal axis, as expected. 

However, even if multiple stationary solutions coexist, one will observe
speckle instability, i.e. a time dependent solution persisting at long
time, only if these solutions are dynamically unstable. To investigate the dynamical behaviour, we
use the following relaxation model for the time evolution of the dipole
$d(t)$:
\begin{equation}
\dot{d}(t)=-\frac{\Gamma}{2} \biggl(d(t)-g(I)E)\biggr),
\end{equation}
where $\Gamma/2$ corresponds to the inverse relaxation time. $E$ and $I$ are now the instantaneous local field and intensity. 
We assume
$\Gamma\ll c/R$, such that the field propagation can be considered
as instantaneous within the radius $R$ of the cloud. According to this model, small deviations 
$\delta\mathbf{E}(t)=\mathbf{E}(t)-\mathbf{E}^{\mathrm{st}}$ from the
stationary solution 
$\mathbf{E}^{\mathrm{st}}$ fulfill the following set of equations:
\begin{equation}
\dot{\delta\mathbf{E}}(t)=-\frac{\Gamma}{2}\mathbf{M}\delta\mathbf{E},
\end{equation}
Hence, the stationary solution is stable if all eigenvalues
of the matrix $M$ have a positive real part. In Fig.~\ref{multisol}
(bottom), we plot the smallest real part of all
eigenvalues along the stationary solution shown on the top. 
%The dashed
%line corresponds to $\mathrm{Re}(\lambda)=0$, i.e. the stability
%threshold. 
Just above $\alpha=0.1$, the solution becomes unstable and
the instability increases with $\alpha$. Furthermore,
in the region of multiple solutions (see around $\alpha=0.5$, for
instance), the different solutions have 
different instability rates. As a closer inspection
reveals, the branch between the two turning points is the most
unstable one, exactly like in the standard bistability scheme~\cite{boyd}. However,
whereas in the usual scheme the two other branches would correspond 
to stable solutions, leading to a bistable behavior, they are, 
in the present case, also unstable.

\begin{figure}
\centerline{\includegraphics[width=8cm]{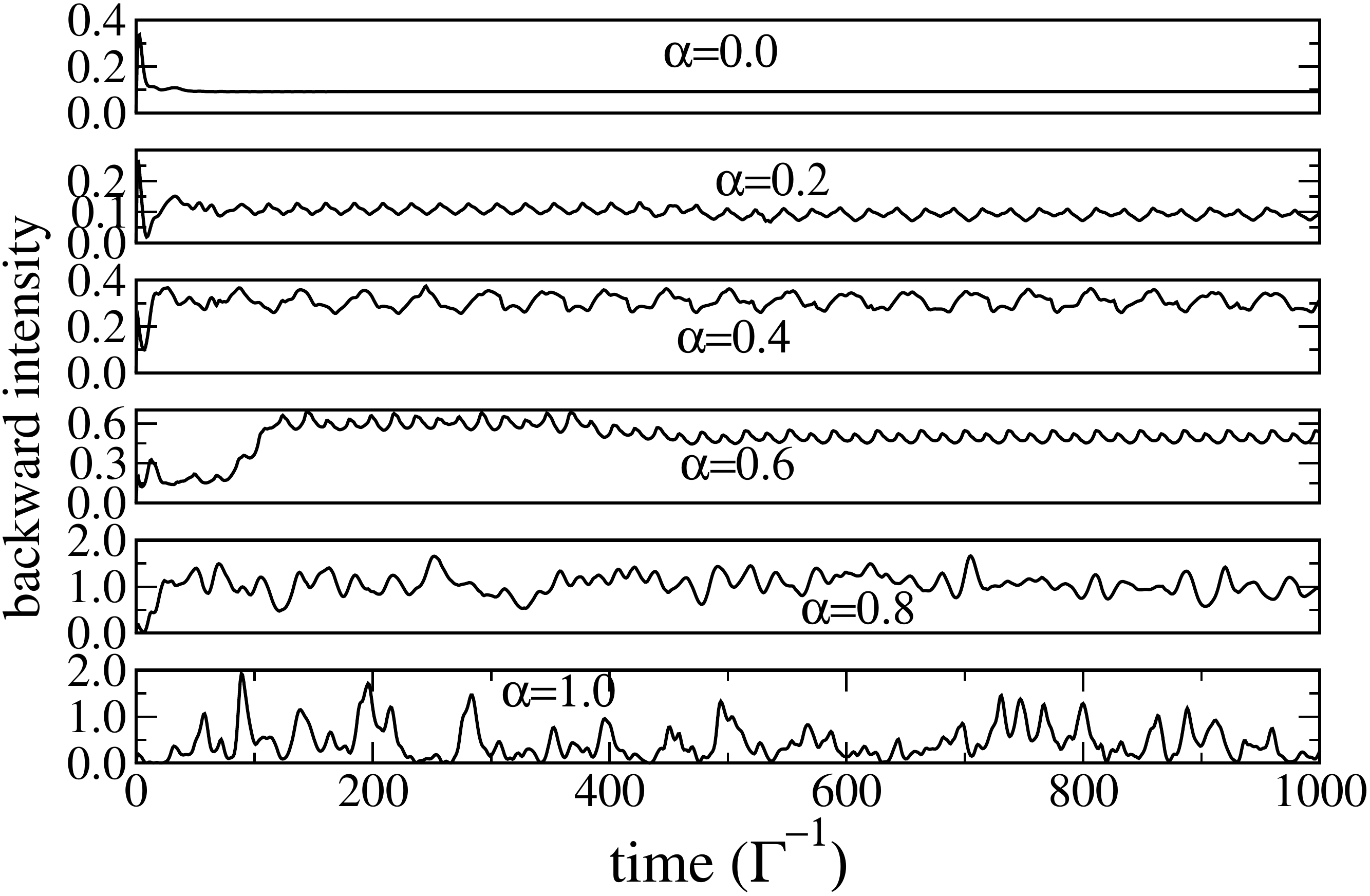}}
\caption{\label{timedep}  Typical evolution of the intensity in the
 backward direction as a function of time for the same configuration
 of scatterers as in Fig.~\ref{multisol}, and for different values
 of the nonlinearity strength $\alpha$. One clearly observes a
transition from a stable stationary solution for $\alpha=0$ to a
chaotic-like behavior for $\alpha=1$. For intermediate values,
stable periodic solutions are obtained, emphasizing that, for those values, the
stationary solutions are already unstable.}
\end{figure}

In summary, Fig.~\ref{multisol}(bottom) reveals that, for the specific configuration of scatterers under consideration, no stable stationary solution exists for
$\alpha>0.1$. Hence, we expect that, in this regime,  the actual dynamics of the
system does not converge towards a stationary state, but remains time-dependent even in the
long-time limit. 
This is confirmed by
Fig.~\ref{timedep}, where we plot, for the same configuration of
scatterers as in Fig.~\ref{multisol}, the backscattered
intensity as a function of time for
different values of the nonlinearity strength $\alpha$. One
clearly observes a
transition from the stable stationary solution at $\alpha=0$ to a
chaotic-like behavior at $\alpha=1$. For intermediate values,
stable periodic solutions are obtained, indicating that, for those values, the
stationary solutions are already unstable. Let us stress
that the bifurcation from the stable to the unstable regime does not
necessarily occur at values of the nonlinearity for which many
stationary solutions coexist. Indeed, since  our
dynamical system has a very large number of
degrees of freedom ($2N$) and, in addition,
is not a Hamiltonian one,  many possible bifurcation scenarios  are in principle possible, including for example strange
attractors~\cite{rasband}. A more detailed analysis would 
require the complete determination of the underlying dynamical structure
(periodic orbits, limit cycles...), which is beyond the scope of this
Letter.   

\begin{figure}
\centerline{\includegraphics[width=9cm]{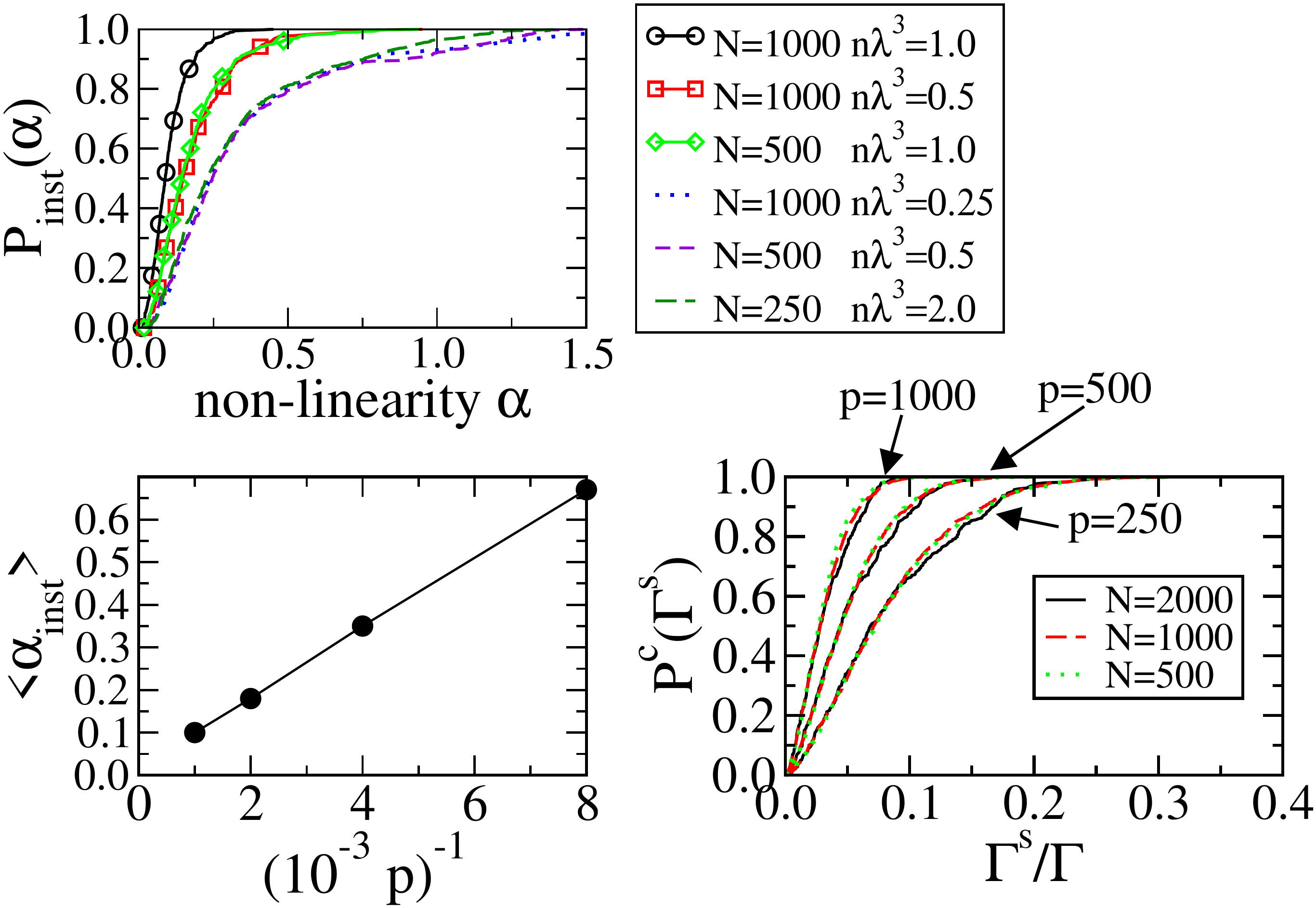}}
\caption{\label{instab_stat} (Color online) On the top, the cumulative distributions of
instability thresholds are plotted for different numbers $N$ and densities $n\lambda^3$ of
scatterers. A scaling law is revealed, according to which the distributions depend only on
a single parameter $p=N\times n\lambda^3$. 
The bottom left plot shows the average threshold to depend linearly on $p^{-1}$.
The bottom right plot displays the distribution $P^c(\Gamma^s)$ of the smallest linear decay rate
$\Gamma_s$, which exhibits the same scaling law in terms of the parameter $p$.
This strongly 
emphasizes the crucial role played by the interference effects for the speckle instability. }
\end{figure}

The preceding discussion refers to a given single configuration of scatterers. Although it
has been chosen as a representative example of the
transition to the unstable regime, more information can be obtained
when performing a statistical analysis of the system properties over an
ensemble of different configurations. 
Fig.~\ref{instab_stat} (top) displays the cumulative distributions of
instability thresholds, denoted $P_{\mathrm{inst}}(\alpha)$, for different system parameters. Hence, we have plotted,
for different numbers $N$ and optical densities
$n\lambda^3$ of scatterers, 
the proportion of unstable configurations as a function of the nonlinearity strength 
$\alpha$. As
expected, if we fix the number of scatterers, 
the instability occurs earlier (i.e. the distributions are shifted towards smaller $\alpha$) for
increasing density (i.e. disorder strength).
On the other hand, if we fix the density, instable configurations are more frequently encountered
with increasing number of scatterers (i.e. system size). 
%Note that the distribution for
%$N=1000$, $n\lambda^3=1$ is consistent with the time evolution
%depicted in Fig.~\ref{timedep}: for $\alpha=0.25$ almost all
%configurations are unstable.
However, both variations are not
independent: as the figure clearly reveals, the statistical
distribution depends only on a single parameter, namely $p=N\times n\lambda^3$.
In particular, the average threshold, $\left\langle\alpha_{\mathrm{inst}}\right\rangle$, depends linearly on $p^{-1}$,
see Fig.~\ref{instab_stat} (bottom, left). 

As we have checked, for large $p$, the resulting 
instability thresholds are small enough so that first order perturbation theory with respect to the
nonlinearity strength $\alpha$ can be applied. Hence, we expect the statistics of
instability thresholds to be
closely related to properties of the linear system, in particular the real parts
of the eigenvalues of $\mathbf M(\alpha=0)$. The latter define the decay rates $\Gamma^s$ of  the linear system \cite{rusek00,pinheiro04}. Since the occurence of a single negative
eigenvalue is already sufficient for the instability to occur, the state with the smallest decay rate
(or, equivalently, longest lifetime) attains a particular importance. 
Therefore, we have plotted in Fig.~\ref{instab_stat}(bottom,
right) the cumulative distribution of the smallest
rate $\Gamma^s$ for different numbers $N$ of scatterers 
and densities $n\lambda^3$. Here, the solid lines correspond
to $N=2000$ and (from left to right) $n\lambda^3=$0.5, 0.25 and 0.125,
the long-dashed lines to $N=1000$ and (from left to right)
$n\lambda^3=$1, 0.5 and 0.25 and the dotted lines to $N=500$ and (from
left to right) $n\lambda^3=$2, 1. and 0.5. And indeed, we 
find the distributions to be governed by the same parameter 
$p=N\times n\lambda^3$ which also determines the statistics of instability thresholds.
From these data, the average value $\Gamma/\left<\Gamma^s\right>$, i.e. the longest lifetime, is found, as a crude estimate,
to scale like $p^{2/3}$. This time is much longer than the
Thouless time, i.e the longest time predicted within diffusion approximation, which, in $\Gamma^{-1}$ unit,
scales like $(p n\lambda^3)^{2/3}$ (i.e. the square of the optical thickness). This difference
strongly emphasizes that interference effects, which enhance the lifetime of a particular single  mode with 
respect to the diffusive Thouless time, are deeply involved in the speckle instability.
 Due to its long lifetime, such a  "prelocalized" mode~\cite{apalkov02} almost behaves like a perfect cavity, and is therefore much more
severely affected by the nonlinearity than short-lived modes and, thus, eventually allows instability to settle.
We note that a similar argument was also put forward to explain the appearance of discrete peaks
in the emission spectrum of the coherent random laser \cite{cao}.
 Let us note that the instability threshold of our nonlinear point scatterers system
  differ from the one predicted in the
case of linear scatterers in a $\chi^{(3)}$ nonlinear medium~\cite{skipetrov},
for which no instability should occur if the nonlinearity is small enough for first order
perturbation theory to apply. Further research is necessary in order to characterize the 
relevant properties of nonlinear disordered systems,
leading to different scenarios for the development of speckle instabilities.

\begin{figure}
\centerline{\includegraphics[width=7cm]{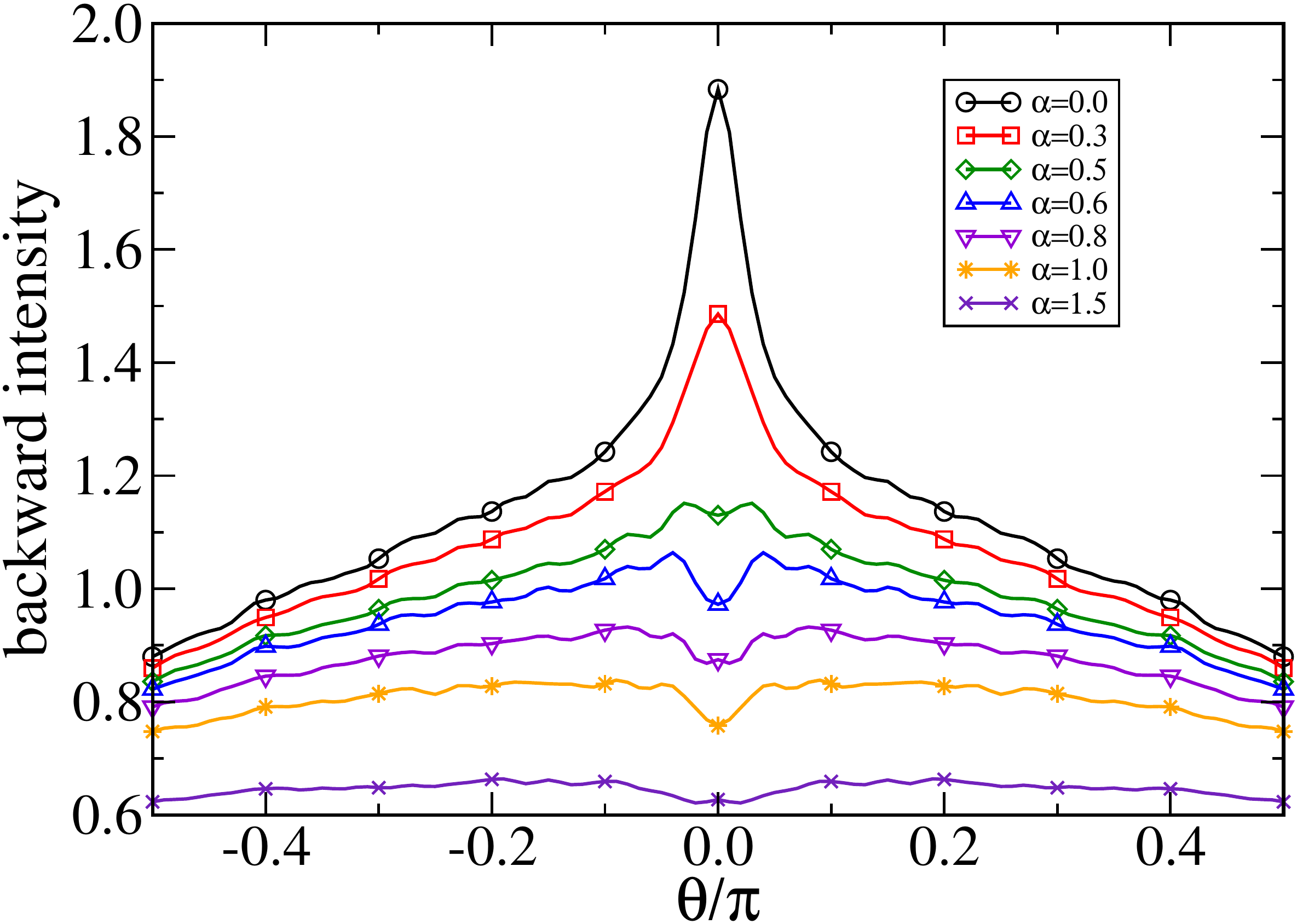}}
\caption{\label{cone_slice} (Color online) Configuration average of
the intensity $\left\langle I(\theta,t)\right\rangle$, scattered at
time $t=1000\Gamma^{-1}$, in the direction $\theta$, where $\theta=0$
corresponds to exact backscattering, for 1000 different configurations of
$N=1000$ scatterers with density $n\lambda^3=1$.
Although the coherent backscattering cone at first decreases with increasing nonlinearity,
the occurence of a dip at larger nonlinearities proves that coherent effects
survive in the time dependent regime $\alpha\gtrsim 0.3$, and even in the fully chaotic
one $\alpha\gtrsim 1$. }
\end{figure}

Finally, we address the question of coherent transport 
in the unstable regime. For this purpose, we compute the coherent backscattering (CBS) cone
$\left\langle I(\theta,t)\right\rangle$, i.e. the configuration average of
the intensity scattered at a time $t$ in the direction $\theta$, where
$\theta=0$ corresponds to the exact backward direction. Numerically,
we observe that after, typically, few $100\Gamma^{-1}$, the average intensity
$\left\langle I(\theta,t)\right\rangle$ becomes time-independent. We plot this stationary value
as a function of $\theta$ in Fig.~\ref{cone_slice}, where the average is performed over 1000 random configurations
with the same parameters ($N=1000$ and $n\lambda^3=1$) as for the configuration
examined in Figs.~\ref{multisol} and \ref{timedep}.
% resulting in a linear optical thickness $b\approx 4$.
Obviously, the net effect of the nonlinearity is the reduction of the
CBS cone height, and the formation of a dip at larger nonlinearities~\cite{hartung08}. According to the theory presented in
\cite{wellens08}, this effect can be qualitatively explained by a dephasing 
between reversed scattering  induced by the nonlinearity. 
In the present context, however, the most important
point is that a coherent effect still survives in the time dependent regime
at $\alpha\gtrsim 0.3$, and even in the fully chaotic one at $\alpha\gtrsim 1$.
A possible explanation can be put forward when
considering the time evolution depicted in Fig~\ref{timedep}: the
typical timescale is much larger than the scatterer response time
$\Gamma^{-1}$, such that the intensity pattern is almost fixed for a
photon scattered along short paths inside the medium.  
%To what extent the diagrammatic approach based on
%the stationary solutions of the nonlinear radiative transfer equation can still be applied
%is an open question~\cite{wellens08}.

In summary, using a model of nonlinear point scatterers, we
have analysed the dynamical instability properties of the speckle
patterns, and observed the transition from a regime
with a unique, stable and stationary solution 
towards chaotic-like behavior. In addition, we have shown that the statistical
properties of the instability thresholds
are determined by the same parameter as the smallest decay rates of the linear system, 
putting thus forward the crucial role
played by interference effects for the speckle
instabilities.
%Furthermore, since for small values of $\alpha$, our model reduces to a
%$\chi^{(3)}$ non-linear scatterer, we expect our results to depict the
%universal behavior for speckle instabilities observed with nonlinear
%point scatterers. 
%This may be wrong: things may be different for different \delta_0...
Furthermore, we have shown that, quite surprisingly, the coherent backscattering
effects are not erased in the time dependent regime, even not in the
chaotic one. Obviously, it would be very interesting to perform similar
studies in the strongly localized regime ($k\ell\lesssim
1$). 
%For instance, an important point to understand is the impact of
%the nonlinearity on the transition itself. is it shifted to larger
%disorder strength? 
Although our argument based on the decay rates of the linear system would then suggest the appearance of instabilities 
for any small amount of nonlinearity, our results also leave open the possibility to observe
strong localization even in presence of nonlinearity, depending on the timescale
 on which the instabilities
develop. The authors would like to thank S.~Skipetrov, C.~Miniatura and D.~Delande for fruitful discussions.


\begin{thebibliography}{99}

% \bibitem{altshuler81}B.L.~Altshuler, A.G.~Aronov and B.Z.~Spivak, JETP Lett. \textbf{33}, 94 (1981).
% 
% \bibitem{altshuler82} B.L.~Altshuler, A.G.~Aronov, and D.E.~Khmelnitskii, J.~Phys.~C \textbf{15}, 7367 (1982).
% 
% \bibitem{physrep140} S.~Chakravarty and A.~Schmid, Phys.~Rep. \textbf{140}, 193 (1986).

\bibitem{clement05}D. Cl\'ement \textit{et al.}, Phys. Rev. Lett {\bf 95} 170409 (2005)

\bibitem{fort05}C. Fort \textit{et al.}, Phys. Rev. Lett. {\bf 95}  170410 (2005)

\bibitem{schulte05}T. Schulte \textit{et al.}, Phys. Rev. Lett. {\bf 95} 170411 (2005) 

\bibitem{lsp07} L.~Sanchez-Palencia \textit{et al.}, Phys.~Rev.~Lett. \textbf{98}, 210401 (2007).

\bibitem{paul07}T. Paul, P. Schlagheck, P. Leboeuf, and N. Pavloff, Phys. Rev. Lett. {\bf 98}, 210602 (2007).

\bibitem{robert07} R.~Kuhn, O.~Sigwarth, C.~Miniatura, D.~Delande and C.A.~Müller, 
New~J.~Phys. \textbf{9}, 161 (2007).


\bibitem{skipetrov08} S.E.~Skipetrov, A.~Minguzzi, B.A.~van~Tiggelen, and B.~Shapiro,
Phys.~Rev.~Lett. \textbf{100}, 165301 (2008).

\bibitem{hartung08} M.~Hartung, T.~Wellens, C.A.~Müller, K.~Richter, and P.~Schlagheck, Phys.~Rev.~Lett. \textbf{101}, 020603 (2008).

\bibitem{lsp08} J. Billy \textit{et al.}, Nature \textbf{453}, 891 (2008).

\bibitem{roati08} G. Roati \textit{et. al.}, Nature \textbf{453}, 895 (2008).

\bibitem{skipetrov}S.~Skipetrov and R.~Maynard,
   Phys. Rev. Lett. \textbf{85}, 736 (2000).

\bibitem{altshuler99} I.L.~Aleiner,  B.L.~Altshuler, and M.E.~Gershenson, Waves Random Media \textbf{9}, 201 (1999).

\bibitem{wellens08} T.~Wellens and B.~Gr\'emaud,
Phys.~Rev.~Lett. \textbf{100}, 033902 (2008).

\bibitem{paul05}T.~Paul \textit{et al.}, Phys. Rev. A
 \textbf{72}, 063621 (2005).


\bibitem{spivak}B.~Spivak and A.~Zyuzin, Phys.~Rev.~Lett.
\textbf{84}, 1970 (2000).


\bibitem{boyd}R.~W.~Boyd, \textit{Nonlinear Optics} Academic Press, San Diego,
 (1992).

\bibitem{rasband} S.N.~Rasband,
\textit{Chaotic dynamics of nonlinear systems}, Ed. Wiley (1990).


% \bibitem{shepel}D.L.~Shepelyansky, Phys. Rev. Lett. \textbf{70}, 1787
%  (1993).

\bibitem{rusek00}M.~Rusek, J.~Mostowski, and A.~Or\l owski, Phys.~Rev.`A \textbf{61}, 022704 (2000)

\bibitem{pinheiro04}F. A. Pinheiro, M. Rusek, A. Or\l owski, and B.A.~van~Tiggelen, Phys.~Rev.~E \textbf{69}, 026605 (2004)

\bibitem{apalkov02}V.M.~Apalkov, M.E.~Raikh, and B.~Shapiro, Phys.~Rev.~Lett. \textbf{89}, 016802 (2002).


\bibitem{cao}H.~Cao, Waves Random Media \textbf{13}, R1 (2003).

% \bibitem{lagendijk96}A.~Lagendijk and B.~van~Tiggelen, Phys.~Rep. \textbf{270}, 143 (1996).
% 
% \bibitem{vanderMark} M.B.~van~der~Mark, M.P.~van~Albada, and
%  A.~Lagendijk, Phys.~Rev.~B \textbf{37}, 3575 (1988).







\end{thebibliography}
\end{document}